\documentclass[aps]{revtex4}

\usepackage[dvips]{graphicx}
\usepackage[dvipdfm]{hyperref}

\begin{document}
\title{Quantum critical behavior in the heavy Fermion single crystal Ce(Ni$_{0.935}$Pd$_{0.065}$)$_2$Ge$_2$}

\author{C. H. Wang$^{1,2,3}$, J. M. Lawrence$^1$, A. D. Christianson$^3$, S. Chang$^4$,
K. Gofryk$^2$, E. D. Bauer$^2$, F. Ronning$^2$, J. D. Thompson$^2$,
K. J. McClellan$^2$,  J. A. Rodriguez-Rivera$^{4,5}$, J. W.
Lynn$^4$}

\address{$^1$University of California, Irvine, California 92697\\
  $^2$Los Alamos National Laboratory, Los Alamos, NM 87545\\
  $^3$Neutron Scattering Science Division, Oak Ridge National Laboratory, Oak Ridge, TN, 37831\\
  $^4$NCNR, National Institute of Standards and Technology, Gaithersburg, MD 20899-6102\\
  $^5$Department of Materials Science and Engineering, University of Maryland, College Park,
  MD 20742\\
}

\email{wangc1@ornl.gov}

\begin{abstract}

We have performed magnetic susceptibility, specific heat,
resistivity, and inelastic neutron scattering measurements on a
single crystal of the heavy Fermion compound
Ce(Ni$_{0.935}$Pd$_{0.065}$)$_2$Ge$_2$, which is believed to be
close to a quantum critical point (QCP) at T = 0. At lowest
temperature(1.8-3.5 K), the magnetic susceptibility behaves as
$\chi(T)-\chi (0)$ $\propto$ $T^{-1/6}$ with $\chi (0) = 0.032
\times  10^{-6}$ m$^3$/mole (0.0025 emu/mole). For $T<$ 1 K, the
specific heat can be fit to the formula $\Delta C/T = \gamma_0 -
T^{1/2}$ with $\gamma_0$ of order 700 mJ/mole-K$^2$. The resistivity
behaves as $\rho = \rho_0 + AT^{3/2}$ for temperatures below 2 K.
This low temperature behavior for $\gamma (T)$ and $\rho (T)$ is in
accord with the SCR theory of Moriya and Takimoto\cite{Moriya}. The
inelastic neutron scattering spectra show a broad peak near 1.5 meV
that appears to be independent of $Q$; we interpret this as Kondo
scattering with $T_K =$ 17 K. In addition, the scattering is
enhanced near $Q$=(1/2, 1/2, 0) with maximum scattering at $\Delta
E$ = 0.45 meV; we interpret this as scattering from
antiferromagnetic fluctuations near the antiferromagnetic QCP.

\end{abstract}

\maketitle

In strongly correlated electron systems, a quantum critical point
(QCP) separates an antiferromagnetic (AFM) or ferromagnetic(FM)
state from a nonmagnetic Fermi liquid state at $T$ = 0 K. This QCP
can be tuned by adjusting a control parameter such as doping
parameter $x$, external pressure $P$, or applied magnetic field $H$.
In the vicinity of the QCP, the critical fluctuations are quantum in
nature and induce unique behavior. The nature of this quantum ground
state phase transition poses one of the most significant challenges
in condensed matter physics. Heavy Fermion (HF) compounds are very
good candidates for studying the QCP. These compounds behave as
Fermi Liquids (FL), with large values for the specific heat
coefficient $\gamma = C/T$ and susceptibility $\chi (0)$, and with
the resistivity varying as $\Delta \rho \propto T^2$ at low
temperature. When such systems are tuned close to a QCP,
fluctuations of the nearby magnetically ordered state affect the
thermodynamic behavior and lead to Non-Fermi Liquid (NFL) behavior
such as $\Delta \rho \propto T^{\alpha}$ with $\alpha <$ 2 and $C/T
\propto ln(T_0/T)$ at low temperature\cite{Stewart}.

The approach to a QCP has been attained in HF systems by doping in
the tetragonal compounds Ce(Ru$_{1-x}$Rh$_x$)$_2$Si$_2$
(x=0.03)\cite{Sekine,Kadowaki1} and Ce$_{1-x}$La$_x$Ru$_2$Si$_2$
(x=0.075)\cite{Kambe,Knafo} as well as  in the orthorhombic compound
CeCu$_{6-x}$Au$_x$ (x=0.1)\cite
{CeCu6Stockert,CeCu6Schroder,Schrodernature}, by application of a
magnetic field in YbRh$_2$Si$_2$\cite{YbCuster}, and with no
additional control parameter in the compound
$\beta$-YbAlB$_4$\cite{YbAlB4}.

A conventional spin fluctuation theory has been established to
explain the non-Fermi liquid behavior. This theory, implemented
through a renormalization-group approach\cite{Hertz,Millis} or
through the self-consistent renormalization(SCR)
method\cite{Moriya}, has successfully explained various aspects of
the non-Fermi liquid behavior. However, the experimental results for
some systems do not follow this spin fluctuation theory. A locally
critical phase transition has been invoked to explain the behavior
of a few systems\cite{Si}. Hence, measurements in other systems are
needed to understand the behavior near the QCP in HF materials.

Below 2 K, the tetragonal compound CeNi$_2$Ge$_2$ exhibits behavior
for the resistivity and specific heat that obeys the predictions of
the SCR theory for a 3D AFM QCP\cite{Moriya}: $\Delta \rho \propto
T^{3/2}$ and $C/T \propto \gamma (0) - A\sqrt{T}$ \cite{Steglich,
Aoki}. Inelastic neutron scattering shows two low energy
features\cite{Kadowaki2,Knopp,Fak}. A broad peak at 4 meV that is
only weakly $\textbf{Q}$-dependent corresponds to Kondo scattering
with $T_K=$ 46 K\cite{Frost}. A peak at 0.7 meV that is highly
$\textbf{Q}$-dependent and shows a maximum intensity at
\textbf{Q$_N$}= (1/2 1/2 0) corresponds to scattering from
antiferromagnetic fluctuations. Although this compound is clearly
close to a QCP, when $T$ decreases below 0.4 K or $B$ increases
above 2 T, the system enters the FL state and $C/T$ shows
saturation\cite{Steglich,Aoki,Koerner}. The compound can be brought
closer to the QCP by alloying with Pd. According to the phase
diagram proposed by Knebel et al\cite{Knebel}, the QCP occurs in
Ce(Ni$_{2-x}$Pd$_x$)$_2$Ge$_2$ when $x=$0.065. Fukuhara et al
found\cite{Fukuhara} that when $x=$ 0.10, the ordering wavevector
remains \textbf{Q$_N$}= (1/2 1/2 0); for larger $x$ it changes to
\textbf{Q$_N$}= (1/2 1/2 1/6).

Here we report measurements of the magnetic susceptibility, the
specific heat for a series of magnetic fields, the resistivity, and
the inelastic neutron scattering of a single crystal of
Ce(Ni$_{0.935}$Pd$_{0.065}$)$_2$Ge$_2$ grown with 58-Ni to reduce
the incoherent scattering. For this alloy, we report the inelastic
neutron spectra for the first time.

Single crystals were grown using the Czochralski method. The
magnetization was measured in a commercial superconducting quantum
interference device (SQUID) magnetometer. The specific heat
measurements were performed in a commercial physical properties
measurement system (PPMS). The electrical resistivity was also
measured in the PPMS using the four wire method. The inelastic
neutron experiment were performed on the MACS
spectrometer\cite{MACS} at the NIST center for neutron research
(NCNR).

\begin{figure}[h]
\begin{minipage}{12pc}
\includegraphics[width=14pc]{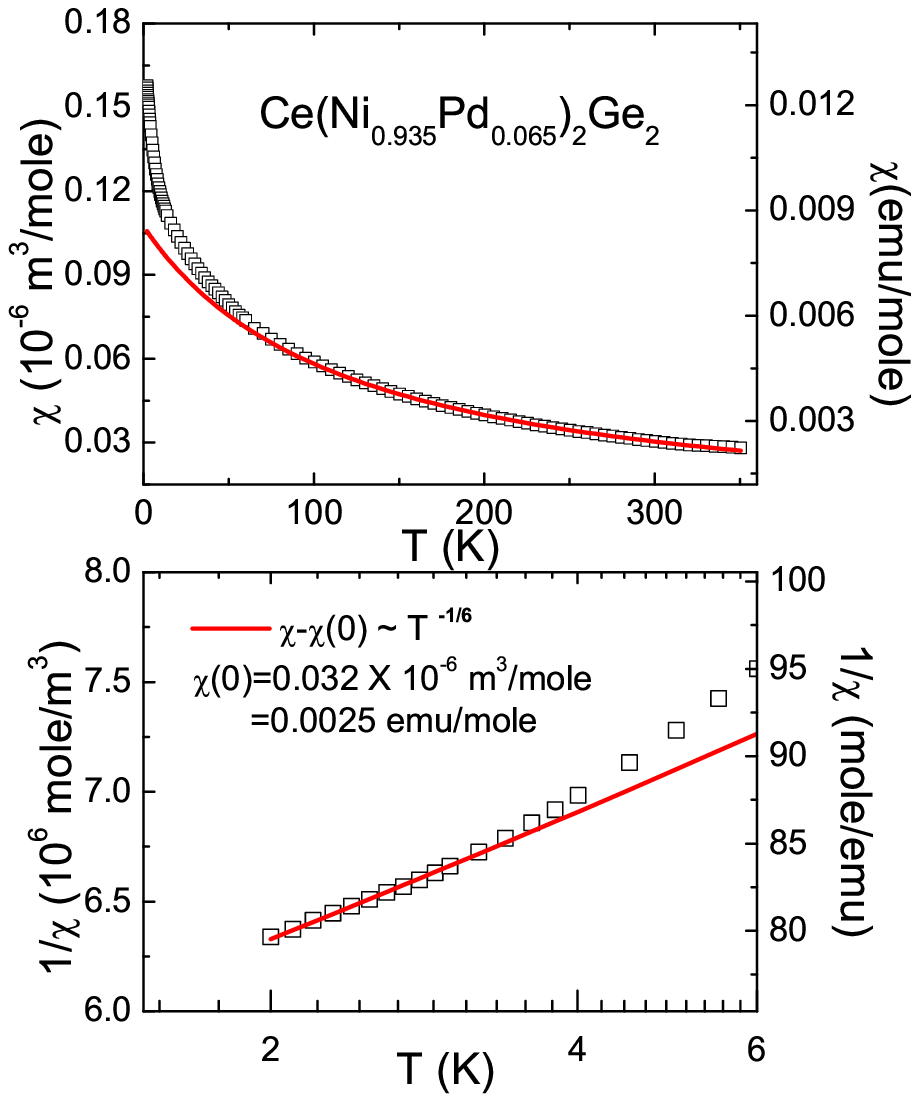}
\caption{\label{label} (a): Magnetic susceptibility of
Ce(Ni$_{0.935}$Pd$_{0.065}$)$_2$Ge$_2$. The solid line is the high
temperature Curie-Weiss fit. (b): Low temperature inverse
susceptibility. The solid line is $\chi(T)- \chi (0) \propto
T^{-1/6}$.}
\end{minipage}\hspace{1pc}%
\begin{minipage}{12pc}
\includegraphics[width=12pc]{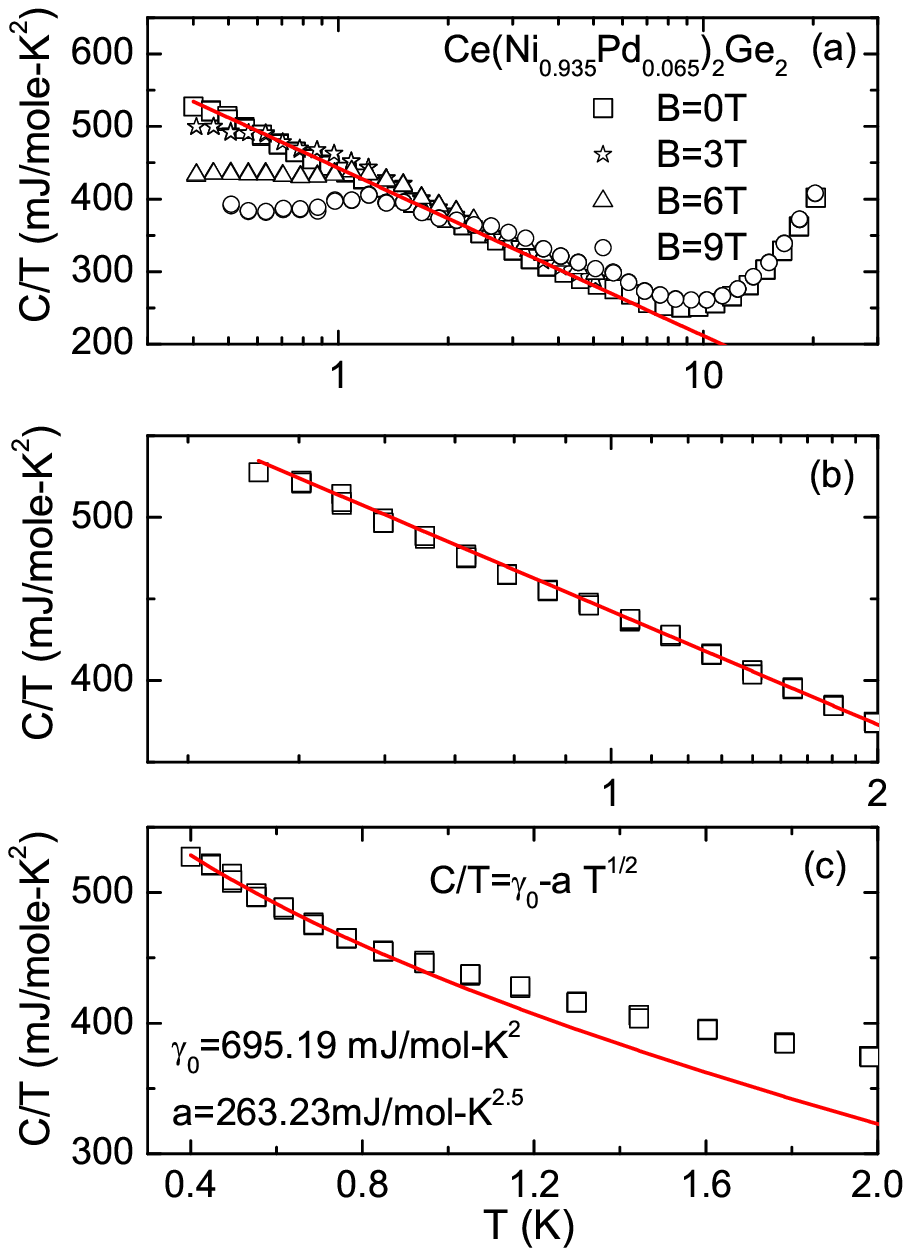}
\caption{\label{label} (a): C/T in applied fields of B=0, 3 T, 6 T
and 9 T. (b): Low temperature zero field C/T curve in a logarithmic
temperature scale. (c) Low temperature zero field C/T data. The
solid line is $\gamma_0-aT^{1/2}$.}
\end{minipage}\hspace{1pc}%
\begin{minipage}{12pc}
\includegraphics[width=12pc]{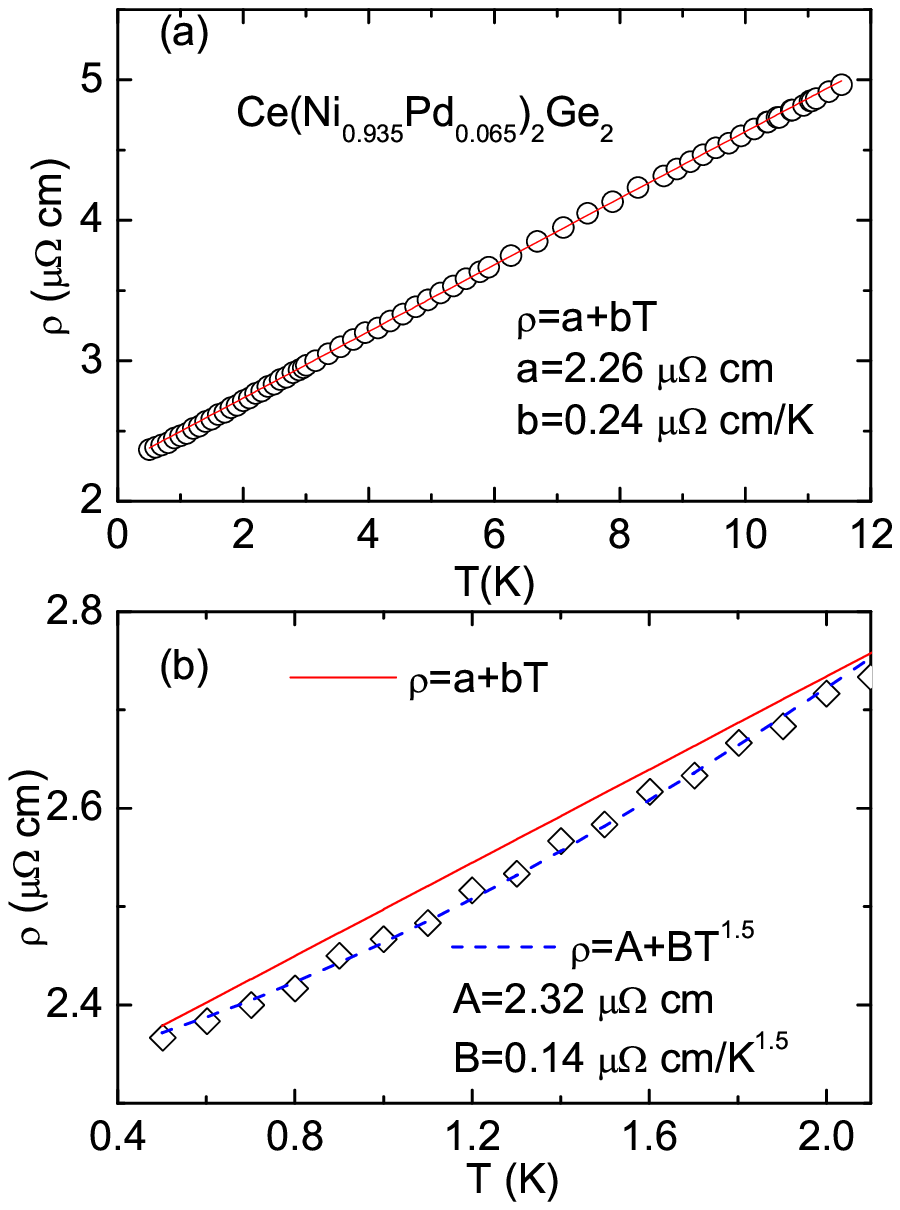}
\caption{\label{label} Resistivity for
Ce(Ni$_{0.935}$Pd$_{0.065}$)$_2$Ge$_2$. The solid line is the linear
fit for the temperature range 0.4 K to 12 K while the dashed line
represents $T^{3/2}$ dependence in the temperature range 0.4 K to 2
K. }
\end{minipage}
\end{figure}

The magnetic susceptibility $\chi(T)$ is shown in figure 1(a). At
high temperatures the data follow Curie-Weiss (CW) behavior $\chi
(T) = C_{eff}/(T- \theta)$; in the range 150-300 K, the CW fit
yields a moment 2.84 $\mu _B$ and $\theta =$ -118 K. Below 100 K,
$\chi(T)$ is enhanced over the CW value but no long range magnetic
order is observed down to 1.8 K. In the traditional spin fluctuation
theory\cite{Moriya,Stewart,Hertz,Millis} for a three dimensional
system near a QCP, $\chi(T) \propto T^{-3/2}$ is expected for an AFM
while $\chi(T) \propto T^{-4/3}$ is expected for a FM. In figure 1
(b), our data for Ce(Ni$_{0.935}$Pd$_{0.065}$)$_2$Ge$_2$ follow the
behavior of $(\chi(T)- \chi (0)) \propto T^{-1/6}$. Exponents for
the susceptibility that are smaller than 1 are also observed for
$\beta$-YbAlB$_4$\cite{YbAlB4} where $\chi(T) \propto T^{-1/3}$ when
B=0.05 T and for YbRh$_2$(Si$_{0.95}$Ge$_{0.05}$)$_2$ where
$(\chi(T)- \chi (0)) \propto T^{-0.6}$\cite{FMYbRh2}.

In figure 2(a) we plot the specific heat $C/T$. The zero field data
are very similar to that reported by Kuwai et al for
Ce(Ni$_{0.936}$Pd$_{0.064}$)$_2$Ge$_2$\cite{Kuwai}. The data are
logarithmic with temperature in the range 1 K to 5 K. When an
external field is applied, the data deviate from the ln$T$ behavior
and show saturation. As for CeNi$_2$Ge$_2$, this is because the
magnetic field forces the system to enter the FL state. However, in
figure 2(b), the low temperature ($T <$1 K) data can be seen to
deviate slightly from the ln$T$ behavior; in this temperature range
the data can be fit(figure 2(c)) to the form $\gamma_0 - a T^{1/2}$
(with $\gamma_0$ of order 695 mJ/mole-K$^2$), which as mentioned is
the expected behavior in the SCR theory.

In figure 3(a) the resistivity $\rho(T)$ data are seen to be roughly
linear in temperature over a wide range 0.4 K to 12 K, indicating
NFL behavior. On the expanded scale of figure 3(b),  $\rho(T)$
deviates from $T$-linear, following a $T^{3/2}$ dependence. Again,
this is the expected behavior in the SCR theory.

Hence, at temperatures higher than 2 K, this alloy exhibits typical
NFL behavior, with the resistivity varying as $\rho(T) \propto T$,
the specific heat varying as $C/T \propto $ln$\frac{T_0}{T}$, and
the susceptibility varying as $(\chi(T)- \chi (0)) \propto
T^{-1/6}$. At temperatures below 1 K, however, $\rho(T) \propto
T^{3/2}$ and $C/T \propto \gamma_0 - a\sqrt{T}$, thereby following
the expected behavior of the SCR theory for a three dimensional AFM
QCP spin fluctuation system.

\begin{figure}[h]
\includegraphics[width=16pc]{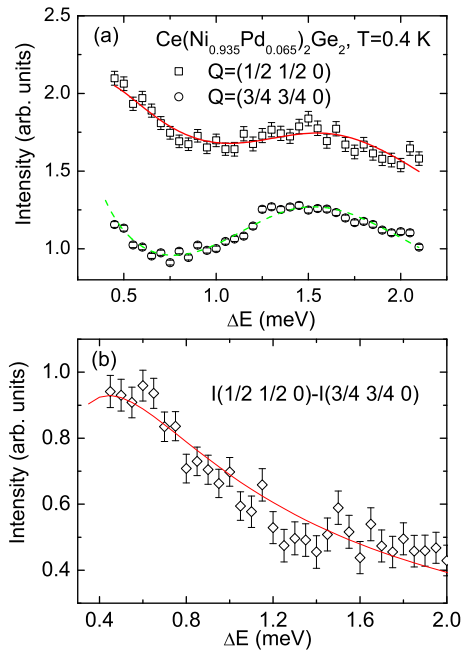}
\begin{minipage}[b]{14pc}\caption{\label{label} Inelastic neutron
scattering spectra for Ce(Ni$_{0.935}$Pd$_{0.065}$)$_2$Ge$_2$. The
data were collected on MACS. In (a), the scattering at the critical
wavevector \textbf{Q$_N$} = (1/2 1/2 0) is compared to the
scattering at (3/4 3/4 0). The lines represent fits to the sum of a
quasi-elastic and an inelastic Lorentzian peak, as described in the
text. In (b) the difference between the data for \textbf{Q}= (1/2
1/2 0) and (3/4 3/4 0) is compared to a quasielastic Lorentzian
(solid line).}
\end{minipage}
\end{figure}

In figure 4, we plot the INS spectra for
Ce(Ni$_{0.936}$Pd$_{0.065}$)$_2$Ge$_2$. A broad peak near 1.5 meV
appears to be only weakly $Q$-dependent. As for the 4 meV peak in
CeNi$_2$Ge$_2$, we identify this as representing Kondo scattering.
Extra intensity is observed  when $\Delta E <$ 1 meV in the vicinity
of $\textbf{Q} =$ (1/2 1/2 0).  This is the wavevector of the
antiferromagnetic fluctuations seen for CeNi$_2$Ge$_2$ near 0.7 meV,
hence we identify it as the critical wavevector \textbf{Q$_N$} for
the QCP.

We first fit these spectra to the sum of a quasi-elastic Lorentzian
(peak width $\Gamma_0$) and an inelastic Lorentzian (peak position
$E_1$ and width $\Gamma_1$) (Fig. 4(a)). For $\textbf{Q} =$ (1/2 1/2
0) we find $E_1$=1.51 meV, $\Gamma_1$=0.90 meV and $\Gamma_0$=0.35
meV; for $\textbf{Q}$=(3/4 3/4 0), we find $E_1$=1.35 meV and
$\Gamma_1$=0.88 meV. In figure 4(b), we use a second approach to
obtain the AFM fluctuation spectra: we subtract the spectra for
$\textbf{Q}=$ (3/4 3/4 0) from that measured at (1/2 1/2 0). A fit
of the difference to a quasielastic Lorentzian peak gives the AFM
fluctuation energy scale as $\Gamma_0 =$ 0.44 meV.

The value of $E_1$ is around 1.35-1.51 meV represents a Kondo
temperature in the range 15.7-17.5 K. We note that this is smaller
than the value 46 K seen in CeNi$_2$Ge$_2$, so that the approach to
the QCP involves a reduction of $T_K$. We note in addition that
$T_K$ does not vanish at the QCP. A finite Kondo temperature is
expected for a spin density wave system at a QCP.

The value of the AFM fluctuation energy $\Gamma_0$ is around
0.35-0.44 meV appears to be smaller than the fluctuation energy
scale of pure CeNi$_2$Ge$_2$ where $\Gamma$=0.7 meV
\cite{Kadowaki2}. This represents "critical slowing down" of the AFM
spin fluctutaions. However, despite the fact that this system is
believed to sit at the QCP, the lifetime of the AFM spin
fluctuations is finite, i.e. it does not diverge as expected at the
critical point. Such a "saturation" of the spin fluctuation lifetime
has been observed for other HF systems where the QCP is attained by
alloying \cite{Kadowaki1, Knafo} and may arise from the fact that
the QCP occurs in a disordered environment.

Research at UC Irvine was supported by the U.S. Department of
Energy, Office of Basic Energy Sciences, Division of Materials
Sciences and Engineering under Award DE-FG02-03ER46036. Work at Los
Alamos National Laboratory was performed under the auspices of the
U.S. DOE/Office of Science. Work at ORNL was sponsored by the
Laboratory Directed Research and Development Program of ORNL,
managed by UT-Battelle, LLC, for the U. S. DOE, and was supported by
the Scientific User Facilities Division, Office of Basic Energy
Sciences, DOE. We acknowledge the support of the National Institute
of Standards and Technology, U. S. Department of Commerce, in
providing the neutron research facilities used in this work. The
identification of  commercial products does not imply endorsement or
recommendation by the National Institute of Standards and
Technology.

\end{document}